# Discrete Cosine-Sine Transform and Fast 16-point Integer Cosine-Sine Transform Low Complexity for Video Coding

L.O. Hnativ*, Ya.V. Luts
V.M. Glushkov Institute of Cybernetics of National Academy of Sciences of Ukraine, Kyiv, Ukraine

**ABSTRACT**

A matrix method for constructing a discrete cosine-sine transform of order $N$ is proposed, upon which discrete and integer cosine-sine transforms of order 16 are constructed. A fast algorithm for computing an integer cosine-sine transform of order-16 with low multiplicative complexity, which is 8.25 times smaller and requires 115% more addition operations compared to the fast discrete cosine-sine transform algorithm, is presented. Additionally, a fast algorithm for computing 2D 16-point integer cosine and cosine-sine separable adaptive transforms with low multiplicative complexity for intra-prediction with 16x16 blocks has been developed. The proposed transforms are effective in the modes of 2D and separable adaptive transforms, providing high coding efficiency in terms of accuracy and compression ratio.

## 1. INTRODUCTION

Transform-based coding methods are used to eliminate redundancy in images and video signals by effectively compressing them. The choice of the best transform for a specific application depends on the value of the permissible recovery error and on the computational cost. In most image and video coding standards, such as JPEG, MPEG -2, H.264/AVC, VC-1, H.265/HEVC [1], use block based transform coding as a tool for efficient compression of input image and video signals. Clark in [2] showed that the optimal Karunen-Loev transform (KLT) for a first-order Markov process with an inseparable correlation model is the discrete cosine transform of type II (DCT-II) [3], when the correlation coefficient $\rho$ between elements approaches unity, and when $\rho$ approaches zero, the corresponding optimal transform is the discrete sine transform (DST) [4]. But with the advent of intra-prediction, this situation changes, and the optimal transform must be adaptive to inter (between frames) or intra-predictive mode. Jain [5,6] showed that the KLT for residual processes will be a Discrete Sine Transform (DST) when the boundary conditions are available in both directions.

Jain [6] called this transform "odd" DST of type 1, and later in [8] it was named "odd" DST of type III (DST-III). A similar result was obtained by Saxena at al. [9,10] for DST type VII (DST-VII). DST-VII is the optimal Karunen-Loev transform for intra-predicted signals. Therefore, DST-VII of order 4 was adopted to the H.265/HEVC standard for intra-residue prediction, which provides better coding characteristics compared to DCT-II. Reznik in [11] the low complexity fast algorithms N-point DST-VII for N=4 and 8 are proposed. In [12,13] shown the connection between DST-VII and DCT-II transforms, the matrix of 2N+1-point DCT-II includes N+1-point DCT-VI and N-point DST-VII. This combination gives fast algorithms for constructing DCT-VI and DST-VII transforms.

In [14], an improved multiple transform ( Enhanced Multiple Tranform, EMT), and received the name adaptive multiple transform (Adaptive Multiple Tranform, AMT), and must be calculated for each block for all difference prediction modes. This leads to a very high computational complexity on the part of the coder [14]. EMT includes 4 transforms DST-VII, DST-I, DCT-V, DCT-VIII. This allows to reduce the bit stream ( BD-Rate) by 3-4%. In work [15] of the new VVC standard (Versatile Video Coding) simplified EMT was named "Multiple Transform selection" (Multiple Transform selection, MTS). In MTS, only DST-VII and DCT-VII remain, excluding other transforms and mode-dependent intra predictions [15].

In [16], fast algorithms with a partial factorization structure of the "artial butterfly" type with an integer DST-VII/DCT-VIII approximation for $N$ = 16, 32, 64 of high computational complexity are proposed, which are used in the VVC video coding standard [15]. Kekre and Solanki [17] introduced a discrete sine-cosine transform of type II, which reduces the average error value by 0.3% compared to DST-II and increases by 0.35% compared to DCT-II and is closest in accuracy to DCT-II compared to other transforms. In [18], a method for constructing an integer modified sine-cosine transform of type VII of order 8 was proposed and two integer transforms were developed with fast algorithms of low multiplicative complexity with high coding efficiency. In [19], a matrix method of discrete cosine-sine transform of type VII of order $N$ is proposed, and two integer cosine-sine transforms of type VII of order 8 with fast algorithms of low multiplicative complexity, which provide high coding efficiency, are built on its basis.

*levhnativ@gmail.com

The paper proposes a matrix method for constructing a discrete cosine-sine transform (DCST) of order $N$. Based on the method, the order-16 discrete cosine-sine transform and integer cosine-sine transform are developed. Fast algorithm for calculation integer cosine-sine transform with low multiplicative complexity compared to the fast algorithm for DCST is developed. A fast algorithm for calculating 2D 16-point integer cosine and cosine-sine separate adaptive transforms of low complexity for intra-prediction with blocks of 16x16 has been developed. Testing of these transforms was carried out and their comparative analysis was made.

## 2. THE METHOD OF CONSTRUCTING A COSINE-SINE TRANSFORM OF ORDER *N*

Let's construct a discrete cosine-sine transform (DCST) matrix of order N, in which rows with even numbers represent the basis cosine functions of the discrete cosine transform of type II ( DCT-II). The odd-numbered rows represent the basis sine functions of the discrete sine transform type IV ( DST-IV), antisymmetric about the middle of the interval. We will call these functions odd modified.

Matrix N - point DCT-II [20] and the N -point DST-IV matrix [21] will be defined as

$$\left[C_N^{II}\right]_{k,n} = \sqrt{\frac{2}{N}}\lambda_k \cos\left[\frac{k(2n+1)\pi}{2N}\right],\ k,n = 0,...,N-1, \tag{1}$$

$$\left[S_N^{IV}\right]_{k,n} = \sqrt{\frac{2}{N}} \sin\left[\frac{(2k+1)(2n+1)\pi}{2N}\right],\ k,n = 0,...,N-1, \tag{2}$$

where the normalization constant $\lambda_k$ is defined as

$$\lambda_k = \begin{cases} 1/\sqrt{2}, k = 0\, or\, k = N, \\ 1 \text{otherwise.} \end{cases}$$

Matrix $CS_N$ N - point DCST will be defined as

$$\left[CS_N\right]_{k,n} = \begin{cases} \sqrt{\frac{2}{N}}\lambda_k \cos\left[\frac{k(2n+1)\pi}{2N}\right], n = 0,...,N-1, k = 0,2,...,N-2, k-even, \\ \sqrt{\frac{2}{N}} \sin\left[\frac{(2k+1)(2n+1)\pi}{2N}\right], n = 0,...,\frac{N}{2}-1, k = 1,3,...,N-1, k-odd, \\ (-1)\sqrt{\frac{2}{N}} \sin\left[\frac{(2k+1)(2n+1)\pi}{2n}\right], n = \frac{N}{2}-1,...,0, k = 1,3,...,N-1. \end{cases} \tag{3}$$

Consider the matrix $DCST_N^*$ size $N \times N$ DCST with permuted rows based on inverse perfect permutations [22]

$$DCST_N^* = P_N DCST_N, \tag{4}$$

where $P_N$ is the matrix $N \times N$ inverse perfect transforms (DP), $P_N(0, N-1) = (0,2,4,...,1,3,...,N-1)$.

Matrix $DCST_N^*$ size $N \times N$ with permuted rows can be constructed using the recursive method:

$$DCST_N^* = 2^{-1/2} diag\left[C_{N/2}^{II}, S_{N/2}^{IV}\bar{I}_{N/2}\right] H_N^*, \tag{5}$$

Where $C_{N/2}^{II}$ is the DCT-II matrix $\frac{N}{2}\times\frac{N}{2}$, $S_{N/2}^{IV}$ is the DST-IV matrix $\frac{N}{2}\times\frac{N}{2}$, $H_N^*$ is the factor matrix $N \times N$ with non-zero elements $\pm 1$,

$$H_N^* = \begin{bmatrix} I_{N/2} & \bar{I}_{N/2} \\ \bar{I}_{N/2} & -I_{N/2} \end{bmatrix}, \bar{I}_{N/2} = antidiag\left[I_{N/2}\right], \bar{I}_{N/2} = \begin{bmatrix} & & & 1 \\ & & 1 & \\ & ... & & \\ 1 & & & \end{bmatrix},$$

$I_{N/2}, \bar{I}_{N/2}$ - unit and antidiagonal unit matrices $\frac{N}{2}\times\frac{N}{2}$.

## 3. THE METHOD OF CONSTRUCTING AN INTEGER COSINE-SINE TRANSFORM OF ORDER-16

Based on method (5), we construct an integer cosine-sine transform (ICST) of order-16. For this, we consider the matrix $ICST_{16}^{*}$ of ICST size $16\times16$ with permuted rows based on binary reverse permutations (BRP), Gray code permutations (GCP) and inverse GCP [22]:

$$ICST_{16}^{*} = \bar{P}_{16}\tilde{P}_{16}ICST_{16}, \tag{6}$$

where $\tilde{P}_{16}$ is a block-diagonal matrix $16\times16$ with matrices $8\times8$ $P_8$ BRP, $\tilde{P}_{16} = diag\left[P_8, P_8\right]$, $P_8\left(\overline{0,7}\right) = \left(0,4,2,6,1,5,3,7\right)$, $\bar{P}_{16}$ - block-diagonal matrix $16\times16$, which contains matrices $4\times4$ $G_4$ GCP and $P_4$ BRP and matrix $8\times8$ $\bar{G}_8$ reverse GCP, $G_4 = diag\left[I_2, \bar{I}_2\right]$, $P_4 = diag\left[1, \bar{I}_2, 1\right]$, $\bar{G}_8\left(\overline{0,7}\right) = \left(0,1,3,2,7,6,4,5\right)$, $\bar{G}_8 = diag\left[G_4, \bar{G}_4\right]$, $\bar{G}_4 = antidiag\left[I_2, \bar{I}_2\right]$, $I_2$, $\bar{I}_2$ - unit and antidiagonal unit matrices $2\times2$, $\bar{I}_2 = \begin{bmatrix} 0 & 1 \\ 1 & 0 \end{bmatrix}$, $\bar{P}_{16} = diag\left[G_4, P_4, \bar{G}_8\right]$.

The matrix $ICST_{16}^{*}$ of the size of $16\times16$ ICT with rearranged rows through the matrix of the ICST core:

$$ICST_{16}^{*} = B_{16}CS_{16}^{*}, \tag{7}$$

where $CS_{16}^{*}$ is the $16\times16$ ICST core matrix with rearranged rows, $B_{16}$ is the diagonal matrix $16\times16$ of normalization coefficients.

The matrix $CS_{16}^{*}$ can be built based on the recurrent method (5):

$$CS_{16}^{*} = diag\left[C_8^{II*}, S_8^{IV*}\bar{I}_8\right]H_{16}^{*}, \tag{8}$$

where $C_8^{II*}$ is the ICT-II core matrix $8\times8$ with permuted rows based on BRP and GCP, $S_8^{IV*}$ - IST-IV core matrix $8\times8$ with permuted rows based on BRP and inverse GCP, $H_{16}^{*}$ - factor matrix $16\times16$ with non-zero elements $\pm1$,

$$H_{16}^{*} = \begin{bmatrix} I_8 & \bar{I}_8 \\ \bar{I}_8 & -I_8 \end{bmatrix}, \bar{I}_8 = antidiag\left[I_8\right], \bar{I}_8 = \begin{bmatrix} & & & 1 \\ & & 1 & \\ & \dots & & \\ 1 & & & \end{bmatrix},$$

$I_8, \bar{I}_8$ - unit and antidiagonal unit matrices $8\times8$.

Matrix $S_8^{IV*}$ IST-IV core of order-8 with permuted rows contains integer elements $\pm a', b', \pm c', d', \pm e', f', \pm g', h', \pm a_i, \pm b_i, \pm c_i$, $i = \overline{1,8}$, and matrix $C_8^{II}$ the ICT-II core of order 8 with rearranged rows contains integer elements $\pm a, \pm b, \pm c, \pm d, \pm e, \pm f, \pm g$, $\pm h, \pm i, \pm j, \pm k$ and have the form:

$$S_8^{IV} = \begin{bmatrix} a' & b' & c' & d' & e' & f' & g' & h' \\ a_1 & a_2 & a_3 & a_4 & a_5 & -a_6 & -a_7 & -a_8 \\ b_1 & b_2 & b_3 & -b_4 & -b_5 & -b_6 & b_7 & b_8 \\ c_1 & c_2 & -c_3 & -c_4 & -c_5 & c_6 & c_7 & -c_8 \\ c_8 & c_7 & -c_6 & -c_5 & c_4 & -c_3 & -c_2 & c_1 \\ b_8 & -b_7 & -b_6 & b_5 & -b_4 & -b_3 & b_2 & -b_1 \\ a_8 & -a_7 & a_6 & a_5 & -a_4 & a_3 & -a_2 & a_1 \\ h' & -g' & f' & -e' & d' & -c' & b' & -a' \end{bmatrix}, \tag{9}$$

where $a' < b' < c' < d' < e' < f' < g' < h'$.

$$C_8^{II} = \begin{bmatrix} k & k & k & k & k & k & k & k \\ a & b & c & d & -d & -c & -b & -a \\ i & j & -j & -i & -i & -j & j & i \\ e & -f & -g & -h & h & g & f & -c \\ k & -k & -k & k & k & -k & -k & k \\ h & -g & f & e & -e & -f & g & -h \\ j & -i & i & -j & -j & i & -i & j \\ d & -c & b & -a & a & -b & c & -d \end{bmatrix}, \qquad (10)$$

where $a > b > c > d$ , $i > j$ , $g > e > h > f$ .

There are relations [24] :

$$\begin{cases} a+d=e \\ a-d=h \end{cases}, \begin{cases} b+c=g \\ b-c=f \end{cases}, \begin{cases} a^2+d^2=b^2+c^2=\alpha \\ e^2+h^2=f^2+g^2=2\alpha \end{cases}$$

Matrix $S_8^{IV*}$ IST-IV core with rearranged rows can be constructed on the basis of the ICT core matrix of order-4 in the form of a product of four matrices:

$$S_8^{IV*} = \tilde{I}_8 H_8^0 diag\left[T_4^*, T_4^*\right] R_8, \qquad (11)$$

where $R_8$ is the rotation-stretch matrix $8\times8$ , which on the main diagonal contains integer elements $\pm S_i$ $\left(i=\overline{1,4}\right)$, and on the opposite diagonal contains integer elements $\pm r_i$, $H_8^0$ - factor matrix $8\times8$ with non-zero elements $\pm1$ , $T_4^*$ - $4\times4$ ICT matrix with rearranged rows based on BRP and GCP, $T_4^* = G_4 P_4 T_4$, $\tilde{I}_8$ - block-diagonal matrix $8\times8$ with unit matrices $4\times4$ $\pm I_4$, $\tilde{I}_8 = diag\left[I_4, -I_4\right]$.

With

$$R_8 = \begin{bmatrix} S_1 & & & & & & & r_1 \\ & S_2 & & & & & r_2 & \\ & & S_3 & & & r_3 & & \\ & & & S_4 & r_4 & & & \\ & & & -r_4 & S_4 & & & \\ & & r_3 & & & -S_3 & & \\ & -r_2 & & & & & S_2 & \\ r_1 & & & & & & & -S_1 \end{bmatrix},$$

$$H_8^0 = \begin{bmatrix} 1 & & & & 0 & & & \\ & 1 & & & & -1 & & \\ & & 1 & & & & 0 & 1 \\ & & & 1 & & & 1 & 0 \\ 0 & & & & 1 & & & \\ & 1 & & & & 1 & & \\ & & 0 & 1 & & & -1 & \\ & & 1 & 0 & & & & -1 \end{bmatrix}$$

where $S_i < S_{i+1}$, $r_i > r_{i+1}$, $i=\overline{1,3}$,

$$T_4^* = \begin{bmatrix} P & P & P & P \\ P & -P & -P & P \\ S & -r & r & -S \\ r & S & -S & -r \end{bmatrix}, \quad r > S, P = 2^n. \qquad (12)$$

For one-norm IST-IV factor matrix $H_8^0$ contains non-zero elements 1 and $\pm P_i / 2^{m_i}$, $i = 1,2,3$, $P_i, m_i$ - integers. Matrix $CS_{16}$ of ICST core of order-16 based on (7)-(10) has the form

$$CS_{16} = \begin{bmatrix} k & k & k & k & k & k & k & k & k & \dots & k \\ a' & b' & c' & d' & e' & f' & g' & h' & -h' & \dots & -a' \\ a & b & c & d & -d & -c & -b & -a & -a & \dots & a \\ a_1 & a_2 & a_3 & a_4 & a_5 & -a_6 & -a_7 & -a_8 & a_8 & \dots & -a_1 \\ i & j & -j & -i & -i & -j & j & i & i & \dots & i \\ b_1 & b_2 & b_3 & -b_4 & -b_5 & -b_6 & b_7 & b_8 & -b_8 & \dots & -b_1 \\ e & -f & -g & -h & h & g & f & -e & -e & \dots & e \\ c_1 & c_2 & -c_3 & -c_4 & -c_5 & c_6 & c_7 & -c_8 & c_8 & \dots & -c_1 \\ k & -k & -k & k & k & -k & -k & k & k & \dots & k \\ c_8 & c_7 & -c_6 & -c_5 & c_4 & -c_3 & -c_2 & c_1 & -c_1 & \dots & -c_8 \\ h & -g & f & e & -e & -f & g & -h & -h & \dots & h \\ b_8 & -b_7 & -b_6 & b_5 & -b_4 & -b_3 & b_2 & -b_1 & b_1 & \dots & -b_8 \\ j & -i & i & -j & -j & i & -i & j & j & \dots & j \\ a_8 & -a_7 & a_6 & a_5 & -a_4 & a_3 & -a_2 & a_1 & -a_1 & \dots & -a_8 \\ d & -c & b & -a & a & -b & c & -d & -d & \dots & d \\ h' & -g' & f' & -e' & d' & -c' & b' & -a' & a' & \dots & -h' \end{bmatrix} \qquad (13)$$

The elements $s_i, r_i$ of the matrix $R_8$ take the following values: $s_1 = 5, s_2 = 14, s_3 = 21,$ $s_4 = 29, r_1 = 45, r_2 = 43, r_3 = 40, r_4 = 35.$ The elements of the matrix $T_4^*$ take the values: $r = 5,\ s = 2,\ p = 2.$ The elements of the matrix $C_8^{II}$ of one-norm scaled ICT-II [23] of order 8 take the following values: $a = 180, b = 152,$ $c = 96, d = 36, e = 152, f = 36, g = 176, h = 100, i = 172$, $j = 56, k = 128.$ The elements of the matrix $S_8^{IV}$ of one-norm scaled IST-IV of order 8 take the following values: $a' = 20, b' = 56, c' = 84, d' = 116, e' = 140, f' = 160, g' = 172, h' = 180.$ The values of the elements $a_i, b_i, c_i\ i = \overline{1,8}$, are given in table 1.

Table 1

| Elements | The values of matrix elements $CS_{16}$ at $i$ | | | | | | | |
|---|---|---|---|---|---|---|---|---|
| | 1 | 2 | 3 | 4 | 5 | 6 | 7 | 8 |
| $a_i$ | 48 | 139 | 180 | 160 | 87 | 18 | 116 | 175 |
| $b_i$ | 85 | 181 | 117 | 55 | 173 | 138 | 11 | 160 |
| $c_i$ | 113 | 161 | 53 | 181 | 16 | 172 | 82 | 141 |

The matrix $S_8^{IV}$ of the proposed scaled IST-IV of order 8 has the form :

$$S_8^{IV} = \begin{pmatrix} 20 & 56 & 84 & 116 & 140 & 160 & 172 & 180 \\ 48 & 139 & 180 & 160 & 87 & -18 & -116 & -175 \\ 85 & 181 & 117 & -55 & -173 & -138 & 11 & 160 \\ 113 & 161 & -53 & -181 & -16 & 172 & 82 & -141 \\ 141 & 82 & -172 & -16 & 181 & -53 & -161 & 113 \\ 160 & -11 & -138 & 173 & -55 & -117 & 181 & -85 \\ 175 & -116 & 18 & 87 & -160 & 180 & -139 & 48 \\ 180 & -172 & 160 & -140 & 116 & -84 & 56 & -20 \end{pmatrix}.$$

The matrix $C_8^{II}$ of the proposed scaled ICT-II of order-8 has the form :

$$C_8^{II} = \begin{pmatrix} 128 & 128 & 128 & 128 & 128 & 128 & 128 & 128 \\ 180 & 152 & 96 & 36 & -36 & -96 & -152 & -180 \\ 172 & 56 & -56 & -172 & -172 & -56 & 56 & 172 \\ 152 & -36 & -176 & -100 & 100 & 176 & 36 & -152 \\ 128 & -128 & -128 & 128 & 128 & -128 & -128 & 128 \\ 100 & -176 & 36 & 152 & -152 & -36 & 176 & -100 \\ 56 & -172 & 172 & -56 & -56 & 172 & -172 & 56 \\ 36 & -96 & 152 & -180 & 180 & -152 & 96 & -36 \end{pmatrix}.$$

The matrix $CS_{16}$ of the proposed scaled ICST of order-16 has the form (14):

$$CS_{16} = \begin{bmatrix} 128 & 128 & 128 & 128 & 128 & 128 & 128 & 128 & 128 & \dots & 128 \\ 20 & 56 & 84 & 116 & 140 & 160 & 172 & 180 & -180 & \dots & -20 \\ 180 & 152 & 96 & 36 & -36 & -96 & -152 & -180 & -180 & \dots & 180 \\ 48 & 139 & 180 & 160 & 87 & -18 & -116 & -175 & 175 & \dots & -48 \\ 172 & 56 & -56 & -172 & -172 & -56 & 56 & 172 & 172 & \dots & 172 \\ 85 & 181 & 117 & -55 & -173 & -138 & 11 & 160 & -160 & \dots & -85 \\ 152 & -36 & -176 & -100 & 100 & 176 & 36 & -152 & -152 & \dots & 152 \\ 113 & 161 & -53 & -181 & -16 & 172 & 82 & -141 & 141 & \dots & -113 \\ 128 & -128 & -128 & 128 & 128 & -128 & -128 & 128 & 128 & \dots & 128 \\ 141 & 82 & -172 & -16 & 181 & -53 & -161 & 113 & -113 & \dots & -141 \\ 100 & -176 & 36 & 152 & -152 & -36 & 176 & -100 & -100 & \dots & 100 \\ 160 & -11 & -138 & 173 & -55 & -117 & 181 & -85 & 85 & \dots & -160 \\ 56 & -172 & 172 & -56 & -56 & 172 & -172 & 56 & 56 & \dots & 56 \\ 175 & -116 & 18 & 87 & -160 & 180 & -139 & 48 & -48 & \dots & -175 \\ 36 & -96 & 152 & -180 & 180 & -152 & 96 & -36 & -36 & \dots & 36 \\ 180 & -172 & 160 & -140 & 116 & -84 & 56 & -20 & 20 & \dots & -180 \end{bmatrix} \qquad (14)$$

The basis vectors of the matrix $CS_{16}$ have a $L_2$ -norm that approaches the power of two: $\|C_i\|^2 = 262144 \pm \Delta i,$ where $\|C_i\|^2$ - the $L_2$-norm $i$ -th of the basis function of the transform matrix (which goes to the unit for the transform with an orthonormal basis), $\Delta i$ (in percent) - the deviation of the $L_2$ -norm of the $i$ -th basis function, which acquires the values $\Delta i = 0{,}12 - 0{,}4\%,\ i = \overline{0{,}15},$ nonorthogonality for cosine functions with even numbers is $0{,}1 - 2\%$ , and for sine functions with odd numbers is $0{,}01 - 0{,}26\%$ .

## 4. ALGORITHM FOR FAST CALCULATION OF 8-POINT DIRECT ICT-II

Matrix $C_8^{II*}$ ICT-II core of order-8 can be factorized as a product of four matrices:

$$C_8^{II*} = C_{8,4}C_{8,3}C_{8,2}C_{8,1}, \tag{15}$$

where $C_{8,i}$ - $i$ -i, $i=\overline{1,4}$, the factor matrix of $8\times 8$ of the fast algorithm proposed in [23] for calculating the 8-point direct ICT-II:

$$\begin{gathered} C_{8,1} = H_8^*, C_{8,2} = diag\left[H_4^*, R_4\right], C_{8,3} = diag\left[T_2, Q_2, H_4\right], C_{8,4} = diag\left[I_4, H_4^0\right], \\ T_2 = \begin{bmatrix} k & k \\ k & -k \end{bmatrix}, Q_2 = \begin{bmatrix} -i & j \\ j & i \end{bmatrix}, H_4^* = \begin{bmatrix} I_2 & \bar{I}_2 \\ \bar{I}_2 & -I_2 \end{bmatrix}, \bar{I}_2 = \begin{bmatrix} 0 & 1 \\ 1 & 0 \end{bmatrix}, \\ H_4^0 = diag\left[1, H_2, 1\right], H_2 = \begin{bmatrix} 1 & 1 \\ 1 & -1 \end{bmatrix}, H_4 = diag\left[H_2, \bar{H}_2\right], \bar{H}_2 = \begin{bmatrix} -1 & 1 \\ 1 & 1 \end{bmatrix}, \end{gathered} \tag{16}$$

$H_4$ - block-diagonal matrix $4\times 4$ with $H_2$ a matrix $2\times 2$ Hadamard and $\bar{H}_2$, $R_4$ - matrix $4\times 4$ rotation-stretch, which on the main diagonal contains integer elements $\pm c$ and $d$ , and on the other diagonal contains integer elements $\pm a$ and $b$ . With

$$H_4 = \begin{bmatrix} 1 & 1 & & \\ 1 & -1 & & \\ & & -1 & 1 \\ & & 1 & 1 \end{bmatrix}, \quad R_4 = \begin{bmatrix} d & & & a \\ & c & b & \\ & b & -c & \\ -a & & & d \end{bmatrix}.$$

For a one-norm ICT-II matrix $H_4^0$ contains non-zero elements 1 and $\pm P/2^m$, $P, m$ - integer numbers.

## 5. ALGORITHM FOR FAST CALCULATION OF 16-POINT DIRECT ICST

Matrix $T_4^*$ of order 4 as the product of two factor matrices:

$$T_4^* = T_{4,2}T_{4,1}, \tag{17}$$

where $T_{4,k}$ - $k$ -th, $k=1,2$, the factor matrix of $4\times 4$ the fast algorithm for calculating the 4-point direct ICT:

$$T_{4,1} = H_4^*, T_{4,2} = diag\left[T_2', R_2\right], T_2' = \begin{bmatrix} P & P \\ P & -P \end{bmatrix}, R_2 = \begin{bmatrix} -r & s \\ s & r \end{bmatrix}. \tag{18}$$

$T_2'$ - a matrix $2\times 2$ with integer elements $\pm P$, $R_2$ - rotation-stretch matrix $2\times 2$ with integer elements $\pm r, s$.

Matrix $CS_{16}^*$ of order-16 according to (8) and taking into account (11), (17) and the fast algorithm for calculating the 8-point direct ICT-II according to (15), (16) can be factorized as a product of six matrices:

$$CS_{16}^* = T_{16,6}, T_{16,5}, T_{16,4}, T_{16,3}, T_{16,2}, T_{16,1}, \tag{19}$$

where $T_{16,k}$ - $k$ -th $k=\overline{1,6}$, matrix factor of $16\times 16$ the proposed fast algorithm for calculating the 16-point direct ICST:

$$\begin{gathered} T_{16,1} = H_{16}^*, T_{16,2} = diag\left[H_8^*, \bar{R}_8\right], T_{16,3} = diag\left[H_4^*, R_4, H_4^*, H_4^*\right], \\ T_{16,4} = diag\left[T_2, Q_2, H_4, T_2', R_2, T_2', R_2\right], T_{16,5} = diag\left[I_4, H_4^0, H_8^0\right], \\ T_{16,6} = diag\left[I_8, I_4, -I_4\right], \bar{R}_8 = R_8\bar{I}_8, \end{gathered} \tag{20}$$

with

$$\bar{R}_8 = \begin{bmatrix} r_1 & & & & & & & s_1 \\ & r_2 & & & & & s_2 & \\ & & r_3 & & & s_3 & & \\ & & & r_4 & s_4 & & & \\ & & & s_4 & -r_4 & & & \\ & & -s_3 & & & r_3 & & \\ & s_2 & & & & & -r_2 & \\ -s_1 & & & & & & & r_1 \end{bmatrix}. \quad (21)$$

## 6. ALGORITHM FOR FAST CALCULATION OF 16-POINT INVERSE ICST

The matrix $CS_{16}^{-1}$ of the inverse ICST core of order-16 can be obtained by transposing:

$$CS_{16}^{-1} = CS_{16}^{*T} / k, \quad (22)$$

where $k = 2^m$, $m$ - integer.

The matrix $CS_{16}^{-1}$ based on (22) and (19)-(21) taking into account the symmetry of the factor matrices $\left(H_N^{*T} = H_N^*, N = 4,8,16, H_4^T = H_4, H_4^{0T} = H_4^0, T_2^T = T_2, T_2'^T = T_2', Q_2^T = Q_2, R_2^T = R_2\right)$ can be factorized by the product of six matrices:

$$CS_{16}^{-1} = T_{16,1} T_{16i,2}^T T_{16i,3}^T T_{16i,4} T_{16,5}^T T_{16,6}, \quad (23)$$

where $T_{16i,k}^T$ - $k$ -th $k = \overline{1,6}$, are the transposed factor matrices $16 \times 16$ of the proposed fast algorithm for calculating the 16-point inverse ICST:

$$\begin{aligned} & T_{16,1} = H_{16}^*, T_{16i,2}^T = diag\left[H_8^*, \bar{R}_{8i}^T\right], T_{16i,3}^T = diag\left[H_4^*, R_{4i}^T, H_4^*, H_4^*\right], \\ & T_{16i,4} = diag\left[T_{2i}, Q_{2i}, H_4, T_{2i}', R_{2i}, T_{2i}', R_{2i}\right], T_{16,5}^T = diag\left[I_4, H_4^0, H_8^{0T}\right], \\ & T_{16,6} = diag\left[I_8, I_4, -I_4\right], \end{aligned} \quad (24)$$

with

$$\bar{R}_{8i}^T = \begin{bmatrix} s_1 & & & & & & & r_1 \\ & s_2 & & & & & r_2 & \\ & & s_3 & & & r_3 & & \\ & & & s_4 & r_4 & & & \\ & & & -r_4 & s_4 & & & \\ & & r_3 & & & -s_3 & & \\ & -r_2 & & & & & s_2 & \\ r_1 & & & & & & & -s_1 \end{bmatrix} / k'. \quad (25)$$

**Implementation of fast calculation of integer transform without multipliers.**

To implement the proposed fast algorithm for calculating 1D 16-point one-norme inverse ICST, mostly shift and addition operations, as well as some multiplication operations, are used. Table 2 presents the scheme for calculating special even multiplications used in the "butterfly" operation to implement the proposed fast algorithm for calculating 1D 16-point inverse ICST.

To implement the proposed fast algorithm for calculating 1D 8-point inverse IST-IV requires 2 multiplication operations, 62 addition operations, and 38 shift operations. Two multiplication operations, 40 addition operations, and 16 shift operations are required to calculate the well-known fast algorithm [23] for calculating 1D 8-point inverse ICT-II.

To implement the proposed fast algorithm for calculating 1D 16-point inverse ICST, 4 multiplication operations, 118 addition operations, and 54 shift operations are required. The algorithm has low multiplicative complexity, which is 8.25 times smaller and requires 115% more addition operations compared to the fast algorithm for discrete cosine-sine transform, developed on the basis of well-known algorithms for DCT-II and DST-IV [25] of order-8.

Table 2

| Multipliers of "butterfly" operations $\overline{R}_{8i}^{T}$, $R_{4i}^{T}$, $Q_{2i}^{T}\left(R_{2i}^{T}\right)$ and matrix elements $H_4^{0T}$, $H_8^{0T}$ | | Algorithm of implementation operations $y = s_i * x;$ $z = r_i * x$ | The number of operations for the implementation of the "butterfly" | | | The number of operations used |
|---|---|---|---|---|---|---|
| | | | Addition | Shift | Multiplication | |
| $S_1 = \frac{5}{32}$ | $r_1 = \frac{45}{32}$ | $x_1 = x \gg 2;\ x_2 = x + x_1;$ $y = x_2 \gg 3;\ z = x_1 + y$ | 2 | 2 | - | 2 |
| $S_2 = \frac{14}{32}$ | $r_2 = \frac{43}{32}$ | $x_1 = x - (x \gg 3);\ y = x_1 \gg 1;$ $x_2 = x + x_1;\ z = x_1 + (x_2 \gg 2)$ | 3 | 3 | - | 2 |
| $S_3 = \frac{21}{32}$ | $r_3 = \frac{40}{32}$ | $x_1 = x \gg 2;\ z = x + x_1;$ $x_2 = z \gg 3;\ y = x_2 + (x \gg 1)$ | 2 | 3 | - | 2 |
| $S_4 = \frac{29}{32}$ | $r_4 = \frac{35}{32}$ | $x_1 = x - (x \gg 2);\ y = x - (x_1 \gg 3);$ $z = x + (x_1 \gg 3)$ | 3 | 2 | - | 2 |
| $s = 2/4$ | $r = \frac{5}{4}$ | $y = x \gg 1;$ $z = x + (x \gg 2)$ | 1 | 2 | - | 4 |
| $P_1 = \frac{181}{256}$ | 0 | $x_1 = 181 * x;$ $y = x_1 \gg 16;\ z = 0$ | - | 1 | 1 | 2 |
| $P_2 = \frac{191}{256}$ | 0 | $x_1 = x - (x \gg 2);\ y = x_1 - (x \gg 16);$ $z = 0$ | 2 | 2 | - | 4 |
| $d = \frac{9}{32}$ | $a = \frac{45}{32}$ | $x_1 = x + (x \gg 3);$ $y = x_1 \gg 2;\ z = x_1 + y$ | 2 | 2 | - | 2 |
| $c = \frac{24}{32}$ | $b = \frac{38}{32}$ | $y = x - (x \gg 2);$ $z = x + (y \gg 2)$ | 2 | 2 | - | 2 |
| $i = \frac{43}{32}$ | $j = \frac{14}{32}$ | $x_1 = x - (x \gg 3);$ $z = x_1 \gg 1;\ x_2 = x + x_1;$ $y = x_1 + (x_2 \gg 2)$ | 3 | 3 | - | 2 |
| $P / 2^m$ | 0 | $x_1 = P * x;\ y = x_1 \gg m;$ $z = 0$ | | 1 | 1 | 2 |
| **That's all** | | - | 42 | 54 | 4 | |

The computational complexity of the proposed fast algorithm for calculating 2D 16-point discrete adaptive ICT-II[23] + ICST, the well-known fast transform DCT-II [25] + DCST is given in the table 3.

Table 3

| **Operations (operations/pixel)** | **Estimation of the computational complexity of 2D discrete separable adaptive transforms for blocks** $16 \times 16$ | | **Comparative analysis** |
|---|---|---|---|
| | Proposed inverse | Proposed | |
| | ICT-II [23] + ICST | DCT-II [25] + DCST | |
| Multiplication | 0.75 | 4, 0625 | In 5, 42 times less |
| Addition + shift | 20.5 | 10, 0625 | 103.73% more additions |

## 7. EXPERIMENTAL RESULTS AND THEIR ANALYSIS

In fig. 1 shows input images of classes B and C for testing.

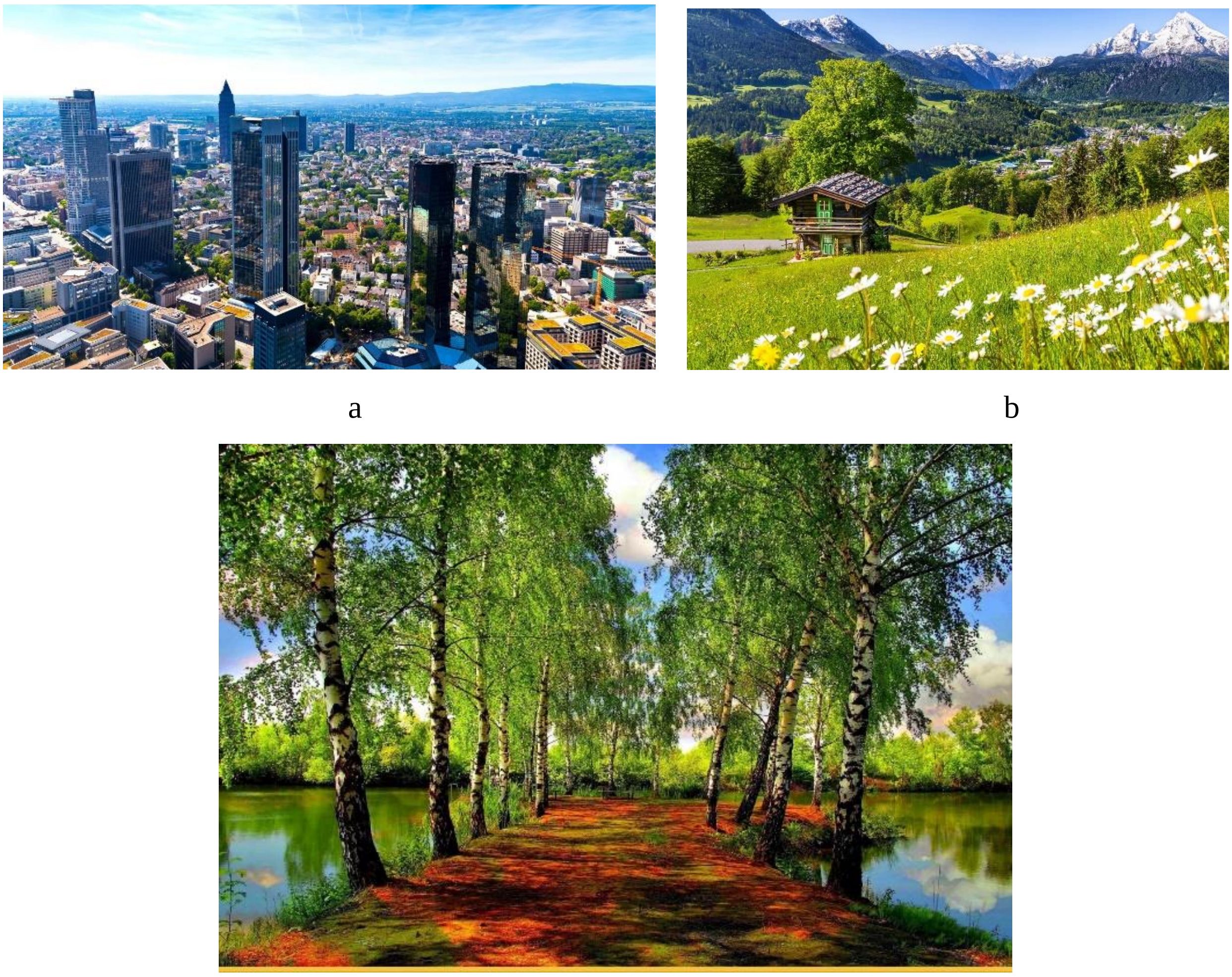

a b

c

Fig. 1. Input images of class B city, size 1920x1072 (a); class C size 1536x1024, landscape (b) and alley (c)

In the table 4 shows the experimental results of coding according to the the standard quantification of PSNR distortions (dB) for compressed test images of class B with a resolution of 1920x1072 pixels (city) (Fig. 1, a), and two images of class C (Fig. 1, b, c) with a resolution of 1536x1024 pixels for the normal and high (22-37) range of QP (quantization parameter) [26] based on 2D proposed DCST and ICST and on the basis of the known DCT-II [25] and the proposed DCST of separate DCT-II /DCST and on the basis of the known ICT-II [20] and the proposed ICST separate ICT-II/ICST.

In the table 5 presents the experimental results for the compression ratio $K$ :1 based on 2D proposed DCST and ICST and separate DCT-II/DCST and ICT-II/ICST. In the table 4 and 5 show the average values of the coding results according to the PSNR characteristics and the compression ratio K for three test images of classes B and C, where LC are low-correlated images.

In the table 6 shows the results of PSNR coding performance based on average values for three test images. These results represent differences based on the proposed ICST and DCST and on the basis of the proposed separate transforms of ICT-II/ICST and separate DCT-II/DCST. In the table 7 shows the results of the coding efficiency by the compression ratio K (in %) based on the proposed ICST and DCST and on the basis of the proposed separate transforms ICT-II/ICST and separate DCT-II/DCST.

During testing, separate adaptive transforms DCT/DCST and ICT/ICST were used without intra-prediction, where DCT and ICT were applied to transform columns and DCST and ICST were applied to transform rows for the purpose of a simplified comparison of two coding performance characteristics (quality and degree compression) of the proposed integer cosine-sine transform with discrete cosine-sine transform.

Table 4

| Image with 16x16 blocks | Proposed Transform (MTS) | **Coding results by PSNR (dB) for QP** | | | | |
|---|---|---|---|---|---|---|
| | | QP=1 | QP =22 | QP = 27 | QP =32 | QP =37 |
| Class B City (LC) 1920x1072 | DCST | 51,395 | 42,447 | 38,276 | 34,105 | 29,656 |
| | ICST | 45,278 | 41,005 | 37,699 | 33,852 | 29,568 |
| | DCT/DCST | 51,397 | 42,543 | 38,438 | 34,316 | 29,913 |
| | ICT /ICST | 45,290 | 41,071 | 37,841 | 34,051 | 29,822 |
| Class C Alley (LC) 1536x1024 | DCST | 51,161 | 41,057 | 36,434 | 32,024 | 27,452 |
| | ICST | 43,146 | 39,232 | 35,720 | 31,744 | 27,354 |
| | DCT/DCST | 51,162 | 41,097 | 36,487 | 32,089 | 27,576 |
| | ICT /ICST | 43,150 | 39,259 | 35,741 | 31,806 | 27.47 7 |
| Class C Landscape (LC) 1536x1024 | DCST | 51,124 | 40,920 | 36,389 | 32,066 | 27,612 |
| | ICST | 44,285 | 39,522 | 35,850 | 31,856 | 27,533 |
| | DCT/DCST | 51,127 | 40,957 | 36,461 | 32,192 | 27,830 |
| | ICT /ICST | 44,291 | 39,546 | 35,914 | 31,974 | 27,750 |
| Average values | DCST | 51,227 | 41,475 | 37,033 | 32,732 | 28,240 |
| | ICST | 44,236 | 39,920 | 36,423 | 32,484 | 28,152 |
| | DCT/DCST | 51,229 | 41,532 | 37,129 | 32,866 | 28,440 |
| | ICT /ICST | 44,244 | 39,959 | 36,499 | 32,610 | 28,350 |

Table 5

| Image with 16x16 blocks | Proposed Transform (MTS) | **Coding results by compression ratio K :1 for QP** | | | | |
|---|---|---|---|---|---|---|
| | | QP=1 | QP =22 | QP = 27 | QP =32 | QP =37 |
| Class B City (LC) 1920x1072 | DCST | 1,437 | 2,225 | 2,722 | 3,583 | 5,916 |
| | ICST | 1,415 | 2,218 | 2,715 | 3,577 | 5,902 |
| | DCT/DCST | 1,456 | 2,342 | 2,888 | 3,821 | 6,382 |
| | ICT /ICST | 1,434 | 2,335 | 2,882 | 3,819 | 6,371 |
| Class C Alley (LC) 1536x1024 | DCST | 1,095 | 1,378 | 1,621 | 2,131 | 3,337 |
| | ICST | 1,087 | 1,378 | 1,617 | 2,126 | 3,328 |
| | DCT/DCST | 1,102 | 1,409 | 1,672 | 2,225 | 3,536 |
| | ICT/ICST | 1,094 | 1,409 | 1,670 | 2,221 | 3,530 |
| Class C Landscape (LC) 1536x1024 | DCST | 1,099 | 1,405 | 1,726 | 2,327 | 3,915 |
| | ICST | 1,094 | 1,403 | 1,723 | 2,323 | 3,904 |
| | DCT/DCST | 1,107 | 1,453 | 1,811 | 2,475 | 4,202 |
| | ICT /ICST | 1,102 | 1,452 | 1,808 | 2,472 | 4,195 |
| Average values | DCST | 1,210 | 1,669 | 2,023 | 2,680 | 4,389 |
| | ICST | 1,199 | 1,666 | 2,018 | 2,675 | 4,378 |
| | DCT/DCST | 1,222 | 1,735 | 2,124 | 2,840 | 4,707 |
| | ICT/ICST | 1,210 | 1,732 | 2,120 | 2,837 | 4,699 |

Table 6

| **Comparison transforms** | **Coding gain results by PSNR (dB) for QP** | | | | |
|---|---|---|---|---|---|
| | QP=1 | QP =22 | QP = 27 | QP =32 | QP =37 |
| ICST with DCST | -6,991 | -1.555 | -0.61 | -0.248 | -0.088 |
| ICT/ICST with DCT/DCST | -6.985 | -1.573 | -0.63 | -0.256 | -0.09 |

Table 7

| Comparison transforms | Coding gain results by compression factor K (in %) for QP | | | | |
|---|---|---|---|---|---|
| | QP=1 | QP =22 | QP = 27 | QP =32 | QP =37 |
| ICST with DCST | -0.909 | -0.180 | -0.247 | -0.187 | -0.251 |
| ICT/ICST with DCT/DCST | -0.982 | -0.173 | -0.188 | -0.106 | -0.170 |

The proposed integer cosine-sine transform of order-16 compared to the discrete cosine-sine transform of order-16 according to the PSNR characteristic at the normal (22, 27) QP range for three test images of classes B and C reduces the average value of PSNR by 1.555 and 0.61DB, for normal (32) and of the high (37) QP range reduces the average value PSNR by 0.248 dB. and 0.088 dB. At the same time, the decrease in the average value of PSNR by 0.248 dB at QP = 32 will be visually imperceptible, according to the subjective threshold of imperceptibility of PSNR < 0.5 dB adopted by the MPEG Committee. That is, the developed transform is effective for the normal (32) and high (37) QP range, and can be adaptively applied in order to reduce the computational complexity. At the same time, the average value of the compression ratio K decreases by 0.187 and 0.0.251%, and for the normal range of QP (22.27), the average value of the compression ratio K decreases by 0.18% and 0.247%.

Analogous, very close results of testing efficiency in terms of PSNR (dB) and compression ratio K (in %) are obtained on the basis of separate transforms of ICT-II/ICST and DCT-II/DCST. According to the results of the experimental analysis, the proposed ICST transform of order-16 is effective both in the 2D mode and in the separate ICT-II/ICST transform mode, as it provides high results of coding efficiency in terms of accuracy and compression ratio. There is a reserve for increasing the coding efficiency due to a more accurate approximation of DCT-II.

## 8. CONCLUSIONS

A matrix method for constructing a discrete cosine-sine transform of order N is proposed, on the basis of which discrete and integer cosine-sine transforms of order-16 are constructed. A fast algorithm for calculating an integer cosine-sine transform of order-16 with low multiplicative complexity, which is 8.25 times smaller and requires 115% more addition operations compared to the fast discrete cosine-sine transform algorithm. A fast algorithm for calculating 2D 16-point integer cosine and cosine-sine separate adaptive transforms of low complexity for intra-prediction with blocks of 16x16 images has been developed. An experimental analysis was carried out, which showed that the proposed transforms are effective in the modes of 2D and separate adaptive transforms, as they provide high coding efficiency in terms of accuracy and compression ratio. There is a reserve for increasing the coding efficiency due to a more accurate approximation of DCT-II.

## REFERENCES

[1] M. Budagavi, A. Fuldseth, G. Bjontegaard, V. Sze and M. Sadafale, "Core Transform Design in the High Efficiency Video Coding (HEVC) Standard", IEEE J. Sel. Topics Signal Process., vol. 7, no. 6, pp. 1029-1041, Dec 2013. DOI: 10.1109/JSTSP.2013.2286118.

[2] Clarke R.J., Relation between the Karhunen-Loeve and Cosine transforms. IEEE Proc. F. Commun., Radar & Signal Process. 1981. Vol. 128. Pt.F, no 6, pp. 359-360. doi: 10.1049/ip-f-1.1981.0061.

[3] N. Ahmed, T. Natarajan and K.R. Rao, “Discrete Cosine Transform,” in IEEE Transactions on Computers, vol. C-23, no. 1, pp. 90-93, Jan. 1974. DOI: 10.1109/TC.1974.223784.

[4] Clarke R.J. Relation between the Karhunen-Loeve and Sine transforms. Electron. Lett., vol. 20, no 1. pp. 12-13. 1984. doi: 10.1049/el:19840010.

[5] Jain A.K. A fast Karhunen-Loeve transform for a class of stochastic processes. IEEE Tran.on Commun., vol. COM-24. no. 9. pp. 1023-1029. 1976.

[6] Jain A.K. Image coding via nearest neighbors image model. IEEE Trans. on Commun., vol. 23, no. 3. pp. 318-321, 1975.

[7] Jain A.K. A sinusoidal family of unitary transforms. IEEE Trans. Pat. Anal. and Mach. Intel., vol. 1, no. 4. pp. 356-365, 1979. DOI: 10.1109/TPAMI.1979.4766944.

[8] Wang Z. and Hunt B.R. The discrete W transform Appl. Math. and Comput., vol. 16, no. 1, pp.. 19-48. 1986. DOI: 10.1016/0096-3003(85)90008-6.
[9] Han J., Saxena A. and Rose K. Towards jointly optimal spatial prediction and adaptive transform in video/image coding. Proc. IEEE Int. Conf. Acoust., Speech, Signal Process (ICASSP), pp. 726-729, March 2010. doi: 10.1109/ICASSP.2010.5495043.
[10] Saxena A. and Fernandes F.C. DCT/DST-based transform coding for intra prediction in image/video coding. IEEE Trans. Image Process., vol. 22, no. 10, pp. 3974–3981, Oct. 2013. doi: 10.1109/TIP.2013.2265882.
[11] Chivukula R.K. and Reznik Y.A. Fast computing of discrete cosine and sine transforms of types VI and VII. In Proc. SPIE 8135, Appl. of Digital Image Processing, XXX IV, no 813505, Sept. 2011, pp. 1-10. doi: 10.1117/12.903685.
[12] Saxena A., Fernandes F.C., and Reznik Y.A. Fast transforms for intra-prediction-based image and video coding, in Proc. 2013 Data Compression Conf., Marth 2013, pp. 13–22. doi: 10.1109/DCC.2013.9.
[13] Reznik Y.A. Relationship between,DCT-II, DCT-VI and DST-VII transforms, in Proc. IEEE Int. Conf. on Acoustics, Speech and Signal Process. (ICASSP) May 2013, pp. 5642–5646. doi: 10.1109/ICASSP.2013.6638744.
[14] T. Biatek, V. Lorcy, P. Castel and P. Philippe, Low-complexity adaptive multiple transforms for post-HEVC video coding, in Picture Coding Symposium, 2016, pp. 1-5. 10.1109/PCS.2016.7906352.
[15] B. Bross, Y.-K. Wang, Y. Ye, S. Liu, GJ Sullivan, and J.-R. Ohm, “Overview of the Versatile Video Coding (VVC) Standard and its Applications,” preprint submitted to IEEE Trans. Circuits and Systs. for Video Tech., December 2020. 10.1109/TCSVT.2021.3060525
[16] Z. Zhang et al., "Fast DST-VII/DCT-VIII With Dual Implementation Support for Versatile Video Coding," in IEEE Transactions on Circuits and Systems for Video Technology, vol. 31, no. 1, pp. 355-371, Jan. 2021. doi: 10.1109/TCSVT.2020.2977118.
[17] Kekre H.B., Solanki J.K. Comparative performance of various trigonometric unitary transforms for transform image coding. Int. J. Electronics, vol. 44, no. 3, pp. 305-315, 1978. doi: 10.1080/00207217808900979.
[18] Hnativ L.O., Luts V.K. Integer Modified Sine-Cosine Transforms Type VII. A construction Method and Separable Directional Adaptive Transforms for Intra Prediction with 8 × 8 Chroma Blocks in Image/Video Coding. Cybern. Syst. Anal. vol. 57, pp.155–164, 2021. https://doi.org/10.1007/s10559-021-00339-9.
[19] Hnativ L.O. Discrete Cosine-Sine Type VII Transform and Fast Integer Transforms for Intra Prediction of Images and Video Coding. Cybern. Syst. Anal. vol. 57, pp. 827–835, 2021. https://doi.org/10.1007/s10559-021-00408-z.
[20] Britanak V., Rao K. R. and Yip P. Discrete Cosine and Sine Transforms: General Properties, Fast Algorithms and Integer Approximations. Oxford. UK: Academic Press. - Elsevier. 2007. DOI: 10.1016/B978-0123736246/50001-4.
[21] Wang Z. Fast algorithms for the discrete W transform and for the discrete Fourier transform. IEEE Trans. Acoust., Speech, Signal Process., vol. 32, no. 8, pp. 803-816, 1984. DOI:10.1109/TASSP.1984.1164450.
[22] B. M. Shevchuk, V. K. Zadiraka, L. O. Hnativ, and S. V. Fraer, Technology of Multifunctional Data Processing and Transmission in Monitoring Networks [in Ukrainian], Naukova Dumka, Kyiv (2010). 375 P. doi: 10.1007/s10559-016-9810-9.
[23] Hnativ L.O. Integer Cosine Transforms for High-Efficiency Image and Video Coding. Cybern. Syst. Anal. vol. 52, pp. 802–816. 2016. https://doi.org/10.1007/s10559-016-9881-7.
[24] Joshi R., Reznik Y., Karczewiz M. Efficient large size transforms for high performance video coding, in Proc. SPIE 7798, Appl. of Digital Image Process. XXXIII, no779831, 2010, pp. 1–7.
[25] Yip P., Rao K.R. Fast decimation-in-time algorithms for a family of discrete sine and cosine transforms. Circuits Systems and Signal Processes, vol. 3, no 4, pp. 387–408. 1984. https://doi.org/10.1007/BF01599167.
[26] Richardson I.E. (2003). H.264 and MPEG-4 Video Compression: Video Coding for Next-generation Multimedia. Chichester: John Wiley & Sons. 349 p. DOI: 10.1002/0470869605.